\documentclass[final,1p,times]{elsarticle}
\usepackage{graphicx,amsmath,amsthm,amssymb,amsfonts,algorithmic,bbm}
\usepackage[bookmarks=false,pdfstartview=FitH,hyperindex=true, colorlinks, linkcolor=black, citecolor=black]{hyperref}
\usepackage{subfigure}
\journal{Nuclear Physics B}
\usepackage[plain]{algorithm}

\theoremstyle{plain}
\newtheorem{theorem}{Theorem}[section]

\newtheorem{lemma}[theorem]{Lemma}

\newtheorem{algthm}{Algorithm}

\theoremstyle{definition}

\theoremstyle{remark}
\newtheorem{remark}{Remark}[section]

\begin{document}
\renewcommand{\algorithmicwhile}{\textbf{randomly with probability}}
\renewcommand{\algorithmicendwhile}{\textbf{end randomized block}}
\newcommand{\RANDOMIZEDBLOCK}{\WHILE}
\newcommand{\ENDRANDOMIZEDBLOCK}{\ENDWHILE}
\newcommand{\indicator}{\mathbf{1}}
\newcommand{\figurewidth}{\columnwidth}
\newcommand{\ontop}[2]{\genfrac{}{}{0pt}{}{#1}{#2}}
\newcommand{\eval}[1]{\left\langle {#1} \right\rangle}
\newcommand{\leval}[1]{\langle {#1} \rangle}
\newcommand{\reval}[1]{\overline{#1}}
\newcommand{\bigast}[1]{\underset{#1}{\textrm{{\huge $\ast$}}}}
\newcommand{\deenne}[2]{\frac{\partial^#2}{\partial #1 ^#2}}
\newcommand{\vett}[1]{#1}
\newcommand{\intmsum} {\left( \int\!\! - \! \sum \right)}
\newcommand{\tinyfrac}[2] {\genfrac{}{}{}{1}{#1}{#2} }
\newcommand{\Lfrac}[2] {\genfrac{}{}{}{0}{#1}{#2} }
\newcommand{\dx}[1] {\mathrm{d}{#1}}
\newcommand{\partlike}[1]{\vspace{4mm} \noindent {\bf #1}}
\newenvironment{provv}{\begin{footnotesize}}{\end{footnotesize}}
\def\reff#1{(\ref{#1})}
\newcommand{\be}{\begin{equation}}
\newcommand{\ee}{\end{equation}}
\newcommand{\<}{\langle}
\renewcommand{\>}{\rangle}
\renewcommand{\emptyset}{\varnothing}
\newcommand{\var}{{\rm var}}
\newcommand{\scra}{{\cal A}}
\newcommand{\scrb}{{\cal B}}
\newcommand{\scrc}{{\cal C}}
\newcommand{\scrd}{{\cal D}}
\newcommand{\scre}{{\cal E}}
\newcommand{\scrf}{{\cal F}}
\newcommand{\scrg}{{\cal G}}
\newcommand{\scrh}{{\cal H}}
\newcommand{\scrk}{{\cal K}}
\newcommand{\scrl}{{\cal L}}
\newcommand{\scrm}{{\cal M}}
\newcommand{\scrn}{{\cal N}}
\newcommand{\scro}{{\cal O}}
\newcommand{\scrp}{{\cal P}}
\newcommand{\scrr}{{\cal R}}
\newcommand{\scrs}{{\cal S}}
\newcommand{\scru}{{\cal U}}
\newcommand{\scrz}{{\cal Z}}
\newcommand{\symdif}{\triangle}
\newcommand{\bfsigma}{\boldsymbol{\sigma}}
\newcommand{\Tr}{{\rm Tr}}
\newcommand{\spin}{\mathbf{s}}
\newcommand{\spincomp}{s}

\begin{frontmatter}
  \title{The O($n$) loop model on a three-dimensional lattice}

  \author[ustc]{Qingquan Liu}
  \ead{liuqq@mail.ustc.edu.cn}
  \author[ustc]{Youjin Deng\corref{cor1}}
  \ead{yjdeng@ustc.edu.cn}
  \author[monash]{Timothy M. Garoni}
  \ead{tim.garoni@monash.edu}
  \author[leiden]{Henk W. J. Bl\"ote}
  \ead{henk@lorentz.leidenuniv.nl}
  \cortext[cor1]{Corresponding author} 
  \address[ustc]{Hefei National Laboratory for Physical Sciences at Microscale,\\ Department of Modern Physics, University of Science and Technology of China, Hefei, 230027, China}
  \address[monash]{School of Mathematical Sciences, Monash University, Clayton, Victoria~3800, Australia}
  \address[leiden]{Instituut Lorentz, Leiden University, P.O. Box 9506, 2300 RA Leiden, The Netherlands}

  \begin{abstract}
    We study a class of loop models, parameterized by a continuously varying loop fugacity $n$, on the {\em hydrogen-peroxide} lattice, which is a three-dimensional cubic lattice of coordination number 3.
    For integer $n>0$, these loop models provide graphical representations for $n$-vector models on the same lattice, while for $n=0$ they reduce to the self-avoiding walk problem.
    We use worm algorithms to perform Monte Carlo studies of the loop model for $n=0,~0.5,~1,~1.5,~2,~3,~4,~5$ and $10$ and obtain the critical points and a number of critical exponents, 
    including the thermal exponent $y_t$, magnetic exponent $y_h$, and loop exponent $y_l$.
    For integer $n$, the estimated values of $y_t$ and $y_h$ are found to agree with existing estimates for the three-dimensional O($n$) universality class.
    The efficiency of the worm algorithms is reflected by the small value of the dynamic exponent $z$, determined from our analysis of the integrated autocorrelation times.
  \end{abstract}

  \date{January 24, 2012}
\end{frontmatter}

\section{Introduction}
\label{introduction}
Among the many model systems studied in the field of statistical mechanics,
two fundamental examples that continue to play a central role are the $n$-vector model~\cite{Stanley68} and $q$-state Potts model~\cite{Potts52,Wu82,Wu84}.
Initially, the parameters $n$ and $q$ can assume only positive integer values.
However, the Kasteleyn-Fortuin mapping~\cite{FortuinKasteleyn72} transforms the Potts model into the random-cluster model~\cite{Grimmett06}, in which $q$ appears as a continuous parameter. 
Likewise, certain $n$-vector spin models on the honeycomb lattice can be mapped~\cite{DomanyMukamelNienhuisSchwimmer81} to nonintersecting loop models which remain well-defined for non-integer $n$.
Both types of mappings integrate out the original spin variables, while newly introduced bond variables define the remaining degrees of freedom.
Each bond configuration can be represented by means of a graph covering a subset of the lattice edges. The transformed partition function specifies the statistical weight of each possible graph.
This defines the probabilistic representation of the transformed model in terms of random geometric objects: clusters \cite{FortuinKasteleyn72} and nonintersecting loops \cite{DomanyMukamelNienhuisSchwimmer81}.
These geometric models play a major role in recent developments of conformal field theory~\cite{DiFrancescoMathieuSenechal97} via their connection with Schramm-Loewner 
evolution (SLE)~\cite{KagerNienhuis04,Cardy05}.

Given a particular lattice, or more generally a graph $G=(V,E)$, we consider the loop model defined for $n,x>0$ by the partition function
\begin{equation}
  Z = \sum_{A} n^{c(A)}\,x^{|A|},
  \label{loop model Z}
\end{equation}
where the sum is over all configurations, $A$, of non-intersecting loops that can be drawn on the edges of $G$,
$|A|$ denotes the number of occupied bonds, and $c(A)$ denotes the number of loops.
It is well known~\cite{DomanyMukamelNienhuisSchwimmer81} that on any graph of maximum degree $3$,
the model \eqref{loop model Z} arises for positive integer $n$ as a loop representation of an $n$-component spin model,
\begin{equation}
  Z = \Tr\,\prod_{ij\in E}\,(1+n\,x\,\spin_i\cdot \spin_j),
  \label{spin model Z}
\end{equation}
where the $\spin_i=(\spincomp^1,\dots,\spincomp^n)\in\mathbb{R}^n$ are unit vectors and $\Tr$ denotes normalized integration with respect to any {\em a priori} measure $\<\cdot\>_0$ on $\mathbb{R}^n$
satisfying $\<\spincomp^{\alpha}\spincomp^{\beta}\>_0=\delta_{\alpha,\beta}/n$ and $\<\spincomp^{\alpha}\>_0=\<\spincomp^{\alpha}\spincomp^{\beta}\spincomp^{\gamma}\>_0=0$. 
In particular, uniform measure on the unit sphere is allowed, which results in an O($n$) symmetric spin model.
In addition, various discrete {\em cubic} measures~\cite{DomanyMukamelNienhuisSchwimmer81} are allowed, in which the spins are constrained to the unit hypercube; see Section~\ref{susceptibility expansion}.
For each of these {\em a priori} measures, the model~\eqref{spin model Z} with $n=1$ is simply the Ising model.

For $n\neq1$, the partition function~\eqref{spin model Z} provides a high-temperature (small $\beta=x\,n$) approximation to the partition function of the $n$-vector model defined by the (reduced) Hamiltonian
\begin{equation}
  {\mathcal H}(\spin_1,\spin_2,\ldots) =
  -\beta\,\sum_{ij\in E}\spin_i \cdot \spin_j.
  \label{n-vector Hamiltonian}
\end{equation}
For spins on the unit sphere, the Hamiltonian~\eqref{n-vector Hamiltonian} with $n=2$ and $n=3$, defines the standard XY and Heisenberg models, respectively,
while its $n\rightarrow \infty$ limit corresponds to the spherical model~\cite{BerlinKac52,Stanley68}.
Since the partition function~\eqref{spin model Z} shares the same symmetry as the Hamiltonian \eqref{n-vector Hamiltonian},
one expects that for a given choice of spin {\em a priori} measure, the phase transitions of the two models will belong to the same universality class.
We note that while~\eqref{loop model Z} has a well-defined probabilistic meaning for all $n,x>0$, the spin model~\eqref{spin model Z} has positive weights only when $x<1/n$.
Finally, we recall~\cite{deGennes79} that the $n\to0$ limit of the $n$-vector model corresponds to the self-avoiding walk (SAW) problem,
however in this case the correspondence is not seen at the level of the partition function, but rather at the level of the two-point functions; see Sections~\ref{Check_worm} and~\ref{susceptibility expansion}.

On the honeycomb lattice, an exact analysis~\cite {Nienhuis82,Nienhuis84,Baxter86,Baxter87} of the model~\eqref{loop model Z} is possible,
which yields the critical point and the universal exponents as a function of $n$ in the range ${-2\le~n\le~2}$.
For three-dimensional lattices, however, comparable exact results are not available, and approximations are necessary.
These exist in the form of renormalization at a fixed number of dimensions \cite{BrezinGuillouZinnJustin76},
the $\epsilon$-expansion \cite{Wilson72}, the $1/n$ expansion \cite{Abe72,Ma72,Suzuki72,FerrellScalapino72}, series expansions \cite{Stanley68} and Monte Carlo simulations \cite{LandauBinder09}.
See~\cite{PelissettoVicari02} and references therein.
In particular, Wolff's embedding algorithm~\cite{Wolff89} is a highly-efficient Monte Carlo method for studying the O($n$) spin model for integer $n>0$.

Loop models with continuously varying $n$ have recently been studied via Monte Carlo simulations~\cite{DengGaroniGuoBloteSokal07,LiuDengGaroni11} in two dimensions,
however, we are unaware of any similar studies in three dimensions.
In~\cite{LiuDengGaroni11}, {\em worm} algorithms were presented for simulating the loop model~\eqref{loop model Z} on any 
3-regular graph\footnote{A $k$-regular graph has precisely $k$ edges incident to each vertex. The descriptions ``$k$-regular'' and ``coordination number $k$'' are therefore synonymous.}
and for any real $n>0$, and these algorithms were used to perform a systematic study of the loop model on the honeycomb lattice as a function of $n$.
In the current article, we use the worm algorithms presented in~\cite{LiuDengGaroni11} to perform an analogous study of the loop model on a three-dimensional 3-regular lattice, 
the so-called {\em hydrogen peroxide} lattice~\cite{Wells77}.
Since the hydrogen-peroxide lattice has coordination number 3, it provides a convenient setting for the study of high-temperature series \cite{LeuBettsElliott69,Leu69}.
Therefore, given that worm algorithms simulate spaces of high-temperature graphs, it is also a natural setting for worm-algorithm studies in three dimensions.

\begin{figure}[htb]
  \begin{center}
    \includegraphics[scale=0.8,trim = 20mm 190mm 110mm 20mm, clip]{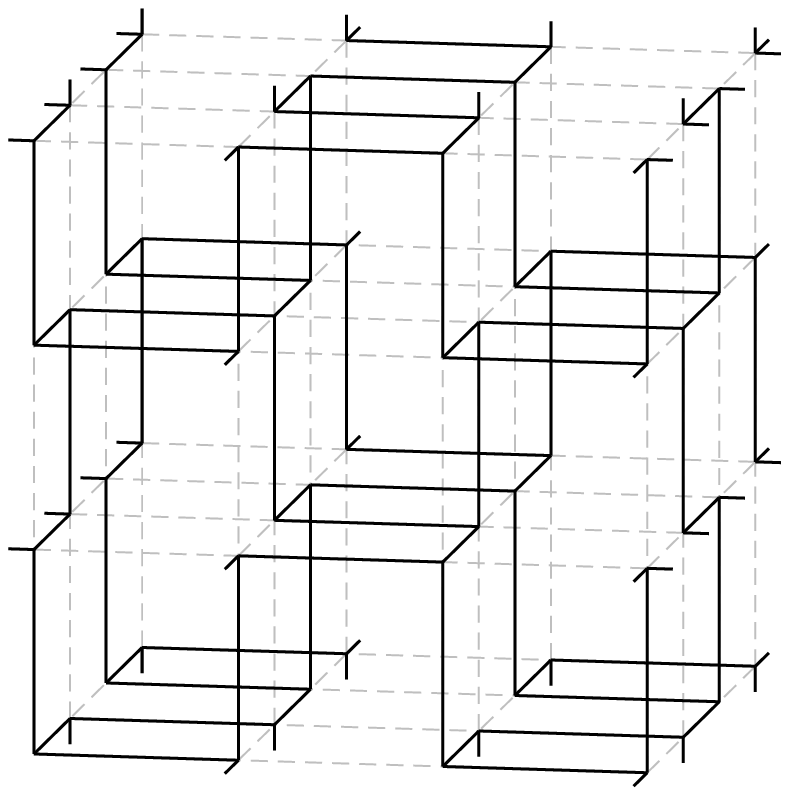}
  \end{center}
  \caption{\label{lattice}
    A 64-site hydrogen peroxide (cubic-cubic) lattice with periodic boundary conditions.
    The sites occupy the same positions as in a $4 \times 4 \times 4$ simple cubic lattice,
    however each site is adjacent to only 3 of the 6 edges allowed in the simple-cubic lattice. 
  }
\end{figure}

For our purposes, the hydrogen-peroxide lattice is perhaps best understood as a subgraph of the simple-cubic lattice.
The vertex set is $\mathbb{Z}^3$, as for the simple-cubic lattice, however precisely half the edges are deleted from the neighborhood of each vertex, so that the edge set is $E_x\cup E_y\cup E_z$ where
\begin{align*}
  E_x &= \{(x,y,z)(x+1,y,z) : x, y, z\in\mathbb{Z},\,\, z + x = 0\mod2\},\\
  E_y &= \{(x,y,z)(x,y+1,z) : x, y, z\in\mathbb{Z},\,\, x + y = 0\mod2\},\\
  E_z &= \{(x,y,z)(x,y,z+1) : x, y, z\in\mathbb{Z},\,\, y + z = 0\mod2\}.
\end{align*}
Since this lattice is a cubic (i.e. $3$-regular) subgraph of the simple-cubic lattice, a more descriptive name may be the ``cubic-cubic'' lattice. 
Being both $3$-regular and transitive, it is in some sense the natural three-dimensional analogue of the honeycomb lattice.
The minimal cycles have length 10, with each vertex belonging to 10 such cycles and each edge to 15.
A sketch of a finite patch wrapped on a 3-dimensional torus is shown in Fig.~\ref{lattice}.
Alternative drawings in which all angles are equal to $120^{\circ}$ can be constructed~\cite{Wells77} and are well-known in crystallography; the international number of such a lattice 
is 214 and its space group is $I4_132$. Both of these geometric configurations possess cubic symmetry.

We shall employ two variants of the worm algorithm in our simulations.
The first version directly utilizes worm transitions that are in detailed balance with the appropriate loop weights $n^{c(A)}\,x^{|A|}$, and is valid for any $n\ge0$.
The $n\to0$ limit of this version is equivalent to the Berretti-Sokal algorithm for SAWs. 
The second version combines the $n=1$ worm algorithm with the coloring method~\cite{DengGaroniGuoBloteSokal07,LiuDengGaroni11} to obtain an alternative algorithm which is valid for real $n\ge1$. 
The advantage of second algorithm is that it avoids the need to perform non-local connectivity queries.

The details of these algorithms will be the subject of Section~\ref{worm algorithms}.
Section~\ref{observables} then describes the observables sampled in our simulations, which include a susceptibility-like quantity, and several dimensionless ratios.
These are used to determine the critical points and the thermal and magnetic exponents by means of finite-size scaling. 
In addition, we present an estimator for the loop fractal dimension $y_l$.
Finally, Section~\ref{results} presents the results of our simulations and our conclusions are summarized in Section~\ref{discussion}.

\section{Worm algorithms}
\label{worm algorithms}
Given a finite graph ${G=(V,E)}$, the cycle space, $\mathcal{C}(G)$, is the set of all $A\subseteq E$ such that every site of $G$ is incident to an even number of bonds in $A$.
We call $A\subseteq E$ and $(V,A)$ {\em Eulerian} whenever $A\in\mathcal{C}(G)$.
The worm algorithm provides a natural method for simulating loop models of the form
\begin{equation}
  Z = \sum_{A\in\mathcal{C}(G)} n^{c(A)}\,x^{|A|},
  \label{combinatorial loop model Z}
\end{equation}
where $c(A)$ is the cyclomatic number of the spanning subgraph $(V,A)$.
The cyclomatic number of a graph is simply the minimum number of edges that need be deleted in order to make it cycle-free.
Consequently, on graphs $G$ of maximum degree 3 all Eulerian configurations $A\in \mathcal{C}(G)$ consist of a collection of $c(A)$ nonintersecting loops, 
and the models \eqref{loop model Z} and \eqref{combinatorial loop model Z} coincide in this case.

The essence of the worm idea is to enlarge the configuration space $\mathcal{C}(G)$ to include two {\em defects} (i.e. vertices of odd degree), and to then move these defects via random walk.
When the two defects happen to occupy the same site they {\em cancel}, and the resulting configuration becomes Eulerian once more.
In the context of spin models, the state space of the worm dynamics corresponds to the space of high-temperature graphs of the two-point correlation function, 
as we describe in more detail in section~\ref{susceptibility expansion}.
Worm algorithms for the high-temperature graphs of classical spin models were first formulated by Prokof'ev and Svistunov~\cite{ProkofevSvistunov01}.
A careful investigation~\cite{DengGaroniSokal07c} of the dynamic behavior of the worm algorithm for the two- and three-dimensional Ising models showed that it suffers from only very weak critical slowing-down,
especially in three dimensions where it was found that the Li-Sokal bound~\cite{LiSokal89} appears to be very close to (or even exactly) sharp.
In Section~\ref{dynamic results} we describe similar results for worm algorithms on the hydrogen peroxide lattice for $n=2$.
This contrasts markedly with the Swendsen-Wang algorithm, for which the Li-Sokal bound is far from sharp in three dimensions~\cite{OssolaSokal04}, 
and even more markedly with algorithms using simple local updates of the ``plaquette'' type~\cite{WinterJankeSchakel08}, which have dynamic exponent $z > 2$.
Furthermore, for certain observables, including the estimator for the susceptibility, the worm algorithm was shown to display critical speeding-up~\cite{DengGaroniSokal07c}.
A discussion of this estimator is the subject of Section~\ref{susceptibility expansion}.

In~\cite{LiuDengGaroni11}, the worm methodology was used to construct algorithms for the loop model~\eqref{combinatorial loop model Z} on any graph $G$ for any real $n>0$, and these algorithms were used to
perform a systematic study of the loop model on the honeycomb lattice. In the current article, we apply the algorithms presented in~\cite{LiuDengGaroni11} to the hydrogen peroxide lattice.
In this section, we give a brief description of these algorithms; refer to~\cite{LiuDengGaroni11} for more details, including a discussion of a version which remains provably ergodic as $x\to\infty$.
Related worm algorithms, corresponding to the loop expansions of both~\eqref{spin model Z} and~\eqref{n-vector Hamiltonian}, are presented in~\cite{Wolff10a,Wolff10c}.

\subsection{Simple worm algorithm}
\label{Check_worm}
It is most natural to describe the worm algorithm on an arbitrary finite graph $G=(V,E)$. For simplicity, we assume $G$ is regular (i.e. each site has the same number of neighbours).
For distinct vertices $u,v\in V$ we let $\mathcal{S}_{u,v}(G)$ denote the set of all subsets of $E$ for which $u$ and $v$ are odd (i.e. have odd degree) with all other vertices being even, and we let
$\mathcal{S}_{v,v}(G)=\mathcal{C}(G)$ for every $v\in V$.
The state space of the worm algorithm is the set $\mathcal{S}(G)$ of all ordered triples $(A,u,v)$ such that $u,v\in V$ and $A\in\mathcal{S}_{u,v}(G)$. 
We emphasize that in the context of spin models $\mathcal{S}_{u,v}(G)$ is nothing other than the set of high-temperature graphs of $\< \spin_u\cdot \spin_v\>$, so that 
$\mathcal{S}(G)$ corresponds to the space of high-temperature graphs of the magnetization; see Section~\ref{susceptibility expansion}.
The proposed moves of the worm algorithm are very simple: starting from $(A,u,v)\in\mathcal{S}(G)$
randomly choose one of the two defects; then randomly choose one of the neighbours of that defect; then move the chosen defect $u$ to the chosen neighbour $u'$ and 
flip the occupation status of the edge $uu'$, so that $A\to A\symdif uu'$. 
Here $A\symdif uu'$ denotes the symmetric difference of $A$ with $uu'$: i.e. delete $uu'$ if it is occupied, or add it if it is vacant.
One then simply applies a standard Metropolis \cite{MetropolisRosenbluthRosenbluthTellerTeller53} acceptance/rejection prescription to determine the
acceptance probabilities required to ensure the worm transitions are in detailed balance with the desired configuration weights $x^{|A|}\,n^{c(A)}$ on $\mathcal{S}(G)$. 
The resulting Monte Carlo algorithm is summarized in Algorithm~\ref{connectivity-checking worm algorithm}. 
We emphasize that for any $(A,u,v)\in\mathcal{S}(G)$, if $u=v$ then $A\in\mathcal{C}(G)$,
so by only measuring observables when $u=v$ one obtains a valid Markov-chain Monte Carlo method for the loop model.
\begin{algorithm}
  \begin{algthm}[Simple worm algorithm] $\,$
    \label{connectivity-checking worm algorithm}
    \begin{algorithmic}
      \LOOP
      \STATE With probability $p$ choose the first defect, otherwise choose the second
      \STATE Uniformly at random, choose a neighbour $u'$ of the chosen defect $u$
      \STATE Propose the update $u\to u'$ and $A\to A\symdif uu'$, leaving the unchosen defect $v$ fixed
      \STATE Accept the proposed update with the acceptance probabilities given in Table~\ref{weights}
      \IF{$u'=v$}
      \STATE Sample observables
      \ENDIF
      \ENDLOOP
    \end{algorithmic}
  \end{algthm}
\end{algorithm}

The explicit acceptance probabilities used in our simulations are listed in Table~\ref{weights}.
Rather than the typical choice of $\min(1,z)$ for the acceptance function we chose $z/(1+z)$, corresponding to a heat-bath algorithm~\cite{SokalLectures}.
We comment on this choice further below.
We emphasize that, in general, the acceptance probabilities depend on the topology of the loops in a non-trivial way; we discuss this issue further below.
The probability $p$ in Algorithm~\ref{connectivity-checking worm algorithm} was set to $1/2$ in our simulations except when $n=0$.
\begin{table}[htb]
  \caption{Acceptance probabilities for the worm proposals $(A,u,v)\to(A\symdif uu',u',v)$. 
    The same acceptance probabilities are used for the proposals ${(A,v,u)\to(A\symdif uu',v,u')}$.}
  \label{weights}
  \begin{tabular}{|r|r|r|r|}
    \hline
    $|A\symdif uu'| - |A|$  & $c(A\symdif uu') - c(A)$ & $\mathbb{P}(A\symdif uu',u',v)/\mathbb{P}(A,u,v)$ & Acceptance Probability \\
    \hline
    $+$1  &  $+$1 &  $nx$        & $nx/(1+nx)$ \\
    $+$1  &   0   &  $x$         & $x/(1+x)$   \\
    $-$1  &  $-$1 &  $1/(nx)$ & $1/(1+nx)$  \\
    $-$1  &   0   &  $1/x$    & $1/(1+x)$   \\
    \hline
  \end{tabular}
\end{table}

The $n\to 0$ limit of Algorithm~\ref{connectivity-checking worm algorithm} deserves some comment.
By construction, the stationary distribution of the worm dynamics is 
\begin{equation}
  \pi_{G,n,x}(A,u,v) = \frac{n^{c(A)}\,x^{|A|}}{\sum_{(A',u',v')\in\mathcal{S}(G)}\,n^{c(A')}\,x^{|A'|}}.
\end{equation}
As $n\to 0$, only those configurations with $c(A)=0$ will retain positive weight, and therefore the support of $\pi_{G,n,x}(\cdot)$ in this limit reduces to the set of all paths (i.e. unrooted SAWs) on $G$.
Likewise, the $n\to0$ limit of Algorithm~\ref{connectivity-checking worm algorithm} provides a valid Monte Carlo algorithm for simulating this reduced space of configurations. 
However, in the context of SAWs, one is generally interested not in the set of {\em all} paths that can be drawn on $G$, but specifically in the subset of such paths which are rooted at a fixed site.
By selecting $p=1$ in Algorithm~\ref{connectivity-checking worm algorithm} however, the location of the second defect becomes fixed and the $p\to1$, $n\to0$ limit of 
Algorithm~\ref{connectivity-checking worm algorithm} defines a valid Monte Carlo procedure for simulating the set of all rooted SAWs on $G$, with fugacity $x$ for the number of steps.
In fact, this algorithm is nothing other than (a slight variation of) the Berretti-Sokal algorithm~\cite{BerrettiSokal85} for the grand-canonical (i.e. variable-length) ensemble of SAWs.
For the simulations reported in Section~\ref{results} therefore, when $n=0$ we used Algorithm~\ref{connectivity-checking worm algorithm} with $p=1$.

An important practical matter when implementing Algorithm~\ref{connectivity-checking worm algorithm} is the need, when $n \neq 0,1$,
to perform a non-local query to determine if the cyclomatic number changes when an update is performed.
Consider a spanning subgraph $(V ,A)\subseteq G$. 
Since the number of components, $k(A)$, is related to the cyclomatic number by $k(A) = |V | - |A| + c(A)$, 
the task of determining whether an edge-update changes the cyclomatic number is equivalent to determining whether it changes the number of connected components. 
The latter question can be answered by known dynamic connectivity-checking algorithms~\cite{HolmDeLichtenbergThorup01}, which take polylogarithmic amortized time. 
A much simpler approach, which runs in polynomial time, but with a (known) small exponent is simultaneous breadth-first search~\cite{DengZhangGaroniSokalSportiello10}.
We used the latter approach when implementing Algorithm~\ref{connectivity-checking worm algorithm} in the simulations presented in Section~\ref{results}. 
In practice however it is worth noting that connectivity queries are not actually required at each step, as we now discuss.

To illustrate, we will consider $n>1$, and suppose the proposed update is to delete an edge $uu'$ which is currently occupied.
In this case the acceptance probability equals either $1/(1+x)$ or $1/(1 +n\,x)$, depending on whether or not $uu'$ is a bridge, and we have the bound $1/(1+n\,x) < 1/(1+x)$ for all $x>0$.
In deciding whether or not to accept the proposed update, we draw a random number $r$ uniformly from $[0,1]$. If $r<1/(1+n\,x)$ then the proposal will be accepted, regardless of whether $uu'$ is a bridge.
Likewise, if $r>1/(1+x)$ then the update will always be rejected. It is therefore only when $1/(1+n\,x) < r < 1/(1+x)$ that we need to perform connectivity queries to determine whether or not $uu'$ is 
a bridge, and therefore whether or not to accept the proposal. A similar observation holds for the proposed addition of an edge, and analogous observations apply also for $n<1$.
Finally, we note also that similar arguments can be used if the acceptance probabilities are chosen using the acceptance function $\min(1,z)$,
however the results are rather more cumbersome in that case. This was our primary motivation for using the heat-bath version.

\begin{remark}
  It is possible to augment Algorithm~\ref{connectivity-checking worm algorithm} with the following additional move: when the two defects collide, move them both to a new vertex of $G$, 
  chosen uniformly at random.
  It was found in~\cite{DengGaroniSokal07c} for the $n=1$ case that the addition of this extra move does not alter the dynamic universality class of the algorithm, 
  but does slightly improve its efficiency, with the efficiency gain tending to zero as the system size tends to infinity.
  For the simulations reported in Section~\ref{results} we applied this extra move when $n>0$.
\end{remark}

\subsection{Coloring algorithm}
\label{Color_worm}
We now describe an alternative to Algorithm~\ref{connectivity-checking worm algorithm} which avoids altogether the need to perform connectivity queries when $n\ge1$. 
Consider a finite graph $G=(V,E)$.
The key observation is that simulating the $n > 1$ loop model on $G$ is equivalent to simulating the $n=1$ loop model on appropriately chosen random subgraphs of $G$.
To see this, let us begin with the loop model partition function~\eqref{combinatorial loop model Z} and introduce auxiliary vertex colorings as follows:
\begin{align}
  Z &= \sum_{A\in\mathcal{C}(G)}\,\prod_{C\in\{\text{cycles of $A$}\}}\,x^{|C|}\,\sum_{\sigma_C\in\{\text{r},\,\text{b}\}}[\delta_{\sigma_C,\,\text{r}} + (n-1)\delta_{\sigma_C,\,\text{b}}],\\
  &= \sum_{A\in\mathcal{C}(G)}\,\sum_{\sigma\in\{\text{r},\,\text{b}\}^V}\,\prod_{ij\in A}\,\delta_{\sigma_i,\sigma_j}\prod_{C\in\{\text{cycles of $A$}\}}\!x^{|C|}\,
  [\delta_{\sigma_C,\,\text{r}} + (n-1)\delta_{\sigma_C,\,\text{b}}]
  \prod_{v\in\{\text{isolated vertices of $A$}\}}\!\!\!\delta_{\sigma_v,\,\text{r}}.
  \label{coloring expansion}
\end{align}
Each vertex can take one of two colors, r(ed) or b(lue), and the leftmost product in~\eqref{coloring expansion} implies that each cycle has all of its vertices colored the same color.
This fact allows us to unambiguously use the notation $\sigma_C$ for the color of a cycle $C$.
The expression~\eqref{coloring expansion} defines a joint model of loops and vertex colorings, and one can simulate the loop model~\eqref{combinatorial loop model Z} by 
simulating the joint model~\eqref{coloring expansion}. This is similar in spirit to the idea behind the Swendsen-Wang cluster algorithm~\cite{SwendsenWang87}.
Given a loop configuration $A\in\mathcal{C}(G)$, the joint model~\eqref{coloring expansion} implies that we can obtain a random vertex coloring by coloring all isolated vertices red,
and independently coloring each cycle red with probability $1/n$ or blue with probability $(n-1)/n$. 
This coloring defines two random induced subgraphs of $G$, one red and one blue, and~\eqref{coloring expansion} implies that on the red subgraph the weight of each loop configuration is
independent of $n$, and is in fact just the weight given by the $n=1$ loop model.
Therefore, the loop configuration on the red subgraph can be updated using an $n=1$ worm algorithm. The weights of loop configurations on the blue subgraph do depend on $n$, but provided we re-choose
the random vertex colorings every so often, to ensure ergodicity, we are free to simply leave the loop configurations on the blue subgraph fixed, since ``do-nothing'' transitions trivially preserve 
stationarity with respect to any measure. We therefore consider the red vertices as being {\em active} and the blue vertices {\em inactive}.
These observations lead to the following algorithm for simulating the $n>1$ loop model~\eqref{combinatorial loop model Z}.
\begin{algthm}[Colored worm algorithm] $\,$
  \label{coloring worm algorithm}
  \begin{algorithmic}
    \LOOP
    \STATE Current state is $A\in\mathcal{C}(G)$
    \RANDOMIZEDBLOCK{$p_{\text{color}}$}
    \STATE Color each isolated vertex red
    \STATE Independently color each loop: red with probability $1/n$, and blue otherwise
    \STATE Identify the induced subgraph of red vertices $G_{\rm red}=G[V_{\rm red}]$
    \ENDRANDOMIZEDBLOCK
    \STATE Choose, uniformly at random, $v\in V(G_{\rm red})$
    \STATE Use $n=1$ worm updates on $G_{\rm red}$ to make a transition $(A_{\rm red},v,v)\to(A_{\rm red}',v',v')$
    \STATE New state is $A'=A_{\rm red}'\cup A_{\rm blue}$
    \ENDLOOP
  \end{algorithmic}
\end{algthm}

The above arguments leading to Algorithm~\ref{coloring worm algorithm} can be expressed in a more formal setting using transition matrices, 
and a formal proof of validity can be given; see~\cite{LiuDengGaroni11}.
The underlying idea behind this {\em coloring method} is in fact very general~\cite{DengGaroniGuoBloteSokal07}.
A similar method was used in~\cite{ChayesMachta98} to construct a cluster algorithm for the random-cluster model for arbitrary real $q>1$.
Related ideas were also discussed in~\cite{Wolff10b}, in the context of a loop model with integer $n$.
We also note that similar ideas were used in an exact mapping between an O($n$) and an O($n-1$) model~\cite{BloteNienhuis89JPhysA}.

The probability $p_{\text{color}}$ in Algorithm~\ref{coloring worm algorithm} can be set to any desired value in $(0,1]$.
  We found empirically that in order to avoid a situation in which almost all of the computer-time is spent on re-coloring, rather than on worm updates, it is advantageous to choose $p_{\text{color}}$ to 
  be strictly less than 1. 
  For each given value of $n>1$ we therefore ran preliminary simulations to tune $p_{\text{color}}$ to a value which resulted in roughly equal time being spent of worm updates and color updates.

  \subsection{Consistency checks}
  We performed several tests to verify the correctness of the worm algorithms.
  First, we verified that Algorithms~\ref{connectivity-checking worm algorithm} and~\ref{coloring worm algorithm} produced the same results for a number of static observables, 
  including the bond density. 
  Furthermore, we checked the consistency of the results of the worm algorithms for general $n$ with an alternative local algorithm which used plaquette updates.
  Some care is needed for these tests because, without special provisions, 
  the plaquette algorithm is subject to conservation laws restricting the numbers of nontrivial loops spanning the periodic system.
  We also performed tests with cluster algorithms for the spin model~(\ref{spin model Z}), for the special cases $n=1$ and $n=2$.
  For the latter case, the tests are restricted to the physical range $x\leq 1/2$, which just excludes the critical point, but still allows accurate tests.

  \section{Observables}
  \label{observables}
  In this section we define the observables that we sampled in our simulations, discuss some quantities of interest defined from these observables, and highlight some of their relevant properties. 
  The numerical results of our simulations and finite-size scaling fits for these quantities are then presented in Section~\ref{results}.

  The following observables were sampled in our simulations.
  All observables were sampled only when the defects coincided, with the exception of the return time, $\mathcal{T}$, which is defined on the full worm chain. The quantity $L$ denotes the linear system size.
  \begin{itemize}
  \item The number of bonds $\mathcal{N}_b(A)=|A|$.
  \item The number of loops $\mathcal{N}_l(A)=c(A)$.
  \item The length of the largest loop $\mathcal{L}_1$.
  \item The mean-square loop length
    \begin{equation}
      \mathcal{L}_2 := L^{-d}\,\sum_{l} |l|^2
      \label{def_quantity_l2}
    \end{equation}
    where the sum is over all loops $l$.

  \item The indicator $\mathcal{W}_x$ for the event that there exists a loop with positive winding number in the $x$ direction. We also sampled $\mathcal{W}_y$ and $\mathcal{W}_z$, which are defined 
    analogously. 
  \item The time $\mathcal{T}$ between consecutive visits to the Eulerian subspace of $\mathcal{S}(G)$ during a realization of the worm chain.
  \end{itemize}
  \bigskip
  From these observables we estimated the following quantities:
  \begin{itemize}
  \item The expectations $\<\mathcal{L}_1\>$ and $\<\mathcal{L}_2\>$.
  \item The dimensionless ratio
    \begin{equation}
      Q_l=\frac{\<\mathcal{L}_1\>^2}{\<\mathcal{L}_1^2\>}.
      \label{Ql definition}
    \end{equation}
    \noindent The quantity $Q_l$ was used to locate the critical point $x_c$ and estimate the thermal exponent $y_t$ when $n>0$.
    It was not studied for $n=0$ since the configurations are single paths rather than multiple loops in this case.
  \item The dimensionless ratio
    \begin{equation}
      Q_b=\frac{\<\mathcal{N}_b\>^2}{\<\mathcal{N}_b^2\>}.
      \label{Qb definition}
    \end{equation}
    \noindent The quantity $Q_b$ was used to locate the critical point $x_c$ and estimate the thermal exponent $y_t$ when $n=0$.
    It was not studied for $n>0$ since $\<\mathcal{N}_b^k\>$ is trivially of order $L^{k\,d}$ for all $x$ in that case.
  \item The wrapping probability
    \begin{equation}
      R := \frac{1}{3}\<\mathcal{W}_x + \mathcal{W}_y + \mathcal{W}_z\>
      \label{wrapping probability definition}
    \end{equation}
    \noindent By symmetry, $R = \<\mathcal{W}_x\> = \<\mathcal{W}_y\> = \<\mathcal{W}_z\>$,
    so the wrapping probability is simply the probability of there existing a loop which has a positive winding number along a fixed coordinate axis.
    Analogous quantities were shown in~\cite{NewmanZiff00} to provide a highly effective method for estimating the critical probability of square-lattice site percolation, and
    various probabilities of this kind can be calculated exactly for the random-cluster model on a square-lattice with toroidal boundary conditions~\cite{Pinson94,Arguin02}.
    A notable feature of the wrapping probabilities studied in~\cite{NewmanZiff00} was their weak finite-size corrections, and the results presented in Section~\ref{results} show that such behavior 
    also holds for the loop model on the hydrogen-peroxide lattice.
  \item The expected time of return $\<\mathcal{T}\>$ to the Eulerian subspace.\\
    \noindent It can be shown for integer $n>0$ that $\<\mathcal{T}\>$ coincides with the susceptibility $\chi$ of the $n$-vector model~\eqref{spin model Z},
    while for $n=0$ it coincides with the susceptibility of the self-avoiding walk problem; see Section~\ref{susceptibility expansion} for details.
    The numerical results presented in~\cite{LiuDengGaroni11} for the honeycomb lattice strongly suggest that in fact $\<\mathcal{T}\>\sim\chi$ holds near criticality
    for arbitrary real $n>0$~\footnote{The case $n=0$ was not studied in~\cite{LiuDengGaroni11}.}, regardless of whether one uses Algorithm~\ref{connectivity-checking worm algorithm} or
    Algorithm~\ref{coloring worm algorithm}.
    Furthermore, the results presented in Section~\ref{results} suggest that $\<\mathcal{T}\>\sim\chi$ also holds on the hydrogen peroxide lattice for real non-negative $n$.
  \end{itemize}
  \bigskip
  We note that the quantities relating to loop properties, $R$, $\<\mathcal{L}_1\>$, $\<\mathcal{L}_2\>$ and $Q_l$, were studied for $n>0$ only.

  \subsection{Worm return times and  susceptibility}
  \label{susceptibility expansion}
  In this section we provide a derivation of the result $\chi=\<\mathcal{T}\>$ for integer $n\ge0$. We consider first the case $n>0$.

  Let $G=(V,E)$ be a finite graph. The two-point correlation function of the model~\eqref{spin model Z} is
  \begin{equation}
    \< \spin_u \cdot \spin_v \> = 
    \frac{1}{Z}\,\Tr \prod_{ij\in E} (1 + n\,x\,\spin_i \cdot \spin_j) \, \spin_u \cdot \spin_v
    \label{correlation function definition}
  \end{equation}
  where the $\spin_i=(\spincomp_i^1,\dots,\spincomp_i^n)\in\mathbb{R}^n$ are unit vectors and $\Tr$ denotes the product over vertices $i\in V$ of normalized integration 
  with respect to some given {\em a priori} $\<\cdot\>_0$ measure on $\mathbb{R}^n$. 
  We will focus on three common cases for the spin {\em a priori} measure, corresponding to the O($n$), corner-cubic and 
  face-cubic models. In each case, the {\em a priori} measure corresponds to uniform measure on a certain compact subset of $\mathbb{R}^n$. In the case of the O($n$) model the spins are constrained to 
  lie on the unit sphere; in the case of the corner-cubic model the spins point to the $2^n$ corners of the unit hypercube; while in the face-cubic model the spins point to the midpoints of the $2n$ faces 
  of the unit hypercube. We note that all three models coincide with the Ising model when $n=1$.
  Furthermore, when $n=2$ the corner-cubic and face-cubic models coincide, since in this case they are simply related by a rotation of $\pi/4$.

  To identify~\cite{DomanyMukamelNienhuisSchwimmer81} the spin partition function~\eqref{spin model Z} with the loop partition function~\eqref{combinatorial loop model Z}, 
  one begins by expanding out the product over edges in~\eqref{spin model Z},
  associating a bond of color $\alpha\in\{1,2,\ldots,n\}$ to each edge $ij$ for which the term $n\,x\,\spincomp_i^{\alpha}\,\spincomp_j^{\alpha}$ is selected.
  One can then attempt to explicitly sum over the spins.
  It is a common feature of each of the above {\em a priori} measures that the moments 
  $\< (s_i^1)^{m_1}\,(s_i^2)^{m_2}\,\ldots\,(s_i^n)^{m_n}\>_0$ vanish unless each $m_{\alpha}$ is even.
  Therefore, only terms for which each vertex is adjacent to an even number of edges of each color can make a non-zero contribution after performing the spin sums.
  One therefore arrives at a model of edge-disjoint colored loop coverings of $G$. 
  For graphs of maximum degree 3, one can then simply sum over all ways of coloring the loops to obtain~\eqref{combinatorial loop model Z}.

\begin{figure}[h]
    \centering
    \subfigure[Tadpole graph.]{
      \includegraphics[scale=1.15,trim = 30mm 250mm 150mm 30mm, clip]{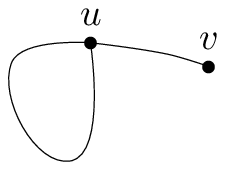}
      \label{tadpole figure}
    }
    \subfigure[Dumbbell graph.]{
      \includegraphics[scale=1.15,trim = 30mm 250mm 145mm 30mm, clip]{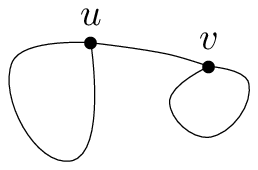}
      \label{dumbbell figure}
    }
    \subfigure[Theta graph.]{
      \includegraphics[scale=1.15,trim = 30mm 250mm 150mm 30mm, clip]{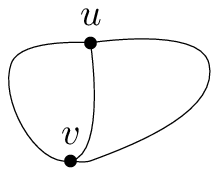}
      \label{theta graph figure}
    }
    \caption{\label{distinct topologies} Possible topologies for the defect cluster. Paths and cycles are also possible, but not shown.}
  \end{figure}

  The same general procedure can be applied to compute a graphical expansion of~\eqref{correlation function definition}.
  In this case, the set of bond configurations leading to non-zero contributions is $\mathcal{S}_{u,v}(G)$,
  the set of all bond configurations with precisely two odd vertices, $u$ and $v$, introduced in Section~\ref{Check_worm}.
  The summation over bond colorings of the component containing $u,v$ (the {\em defect cluster}) must be treated somewhat carefully however, and it turns out that the precise result 
  depends on the specific choice of {\em a priori measure}. In particular, different {\em a priori} measures can give different weights depending on the topology of the defect cluster.
  For graphs of maximum degree 3, the topology of the defect cluster is rather constrained and can be only one of the following:
  cycle, path, tadpole, dumbbell or theta graph. See Fig.~\ref{distinct topologies}.
  In general, whenever the defect cluster is a theta graph it receives an additional model-dependent weight, in addition to $x^{|A|}\,n^{c(A)}$.
  A precise statement of the final result is given in the following lemma, whose detailed proof is straightforward and therefore omitted.
  \begin{lemma}
    {  
      For any graph of maximum degree 3 we have
      $$
      Z\,\< \spin_u\cdot\spin_v\> = \sum_{A\in\mathcal{S}_{u,v}}\,x^{|A|}\,n^{c(A)}\,\Theta(A)
      $$
      where the function $\Theta : \mathcal{S}_{u,v} \to \mathbb{R}$ is
      $$
      \Theta(A) := 
      \begin{cases}
        \theta(A), & \text{if the component containing $u$ and $v$ is a theta graph}, \\
        1,        & \text{otherwise},\\
      \end{cases}
      $$
      and
      $$
      \theta(A) := 
      \begin{cases}
        1,        & \text{face-cubic},\\
        (3n-2)/n^2, & \text{corner-cubic},\\
        3/(n+2), & $O(n)$.\\
      \end{cases}
      $$
      \label{correlation function expansion lemma}
    }
  \end{lemma}

  We note that, unlike the corner-cubic and O($n$) cases, the result for the face-cubic model does not actually involve special weights for the defect cluster, 
  because distinct colors correspond to orthogonal spin states in this case.
  Also, we note that, as expected, the expansion for the corner-cubic model coincides with that for the face-cubic model when $n=1,2$. 
  The O($n$) model only coincides with the face-cubic model for $n=1$.

  Now, from Lemma~\ref{correlation function expansion lemma} we immediately obtain an expansion for the magnetization. 
  By definition $\mathcal{M}=\sum_{i\in V} \spin_i$, so Lemma~\ref{correlation function expansion lemma} implies
  \begin{align}
    \<\mathcal{M}^2\> &= \sum_{u,v\in V}\<\spin_u\cdot\spin_v\>,\\
    &= \frac{1}{Z}\,\sum_{(A,u,v)\in\mathcal{S}(G)}\,x^{|A|}\,n^{c(A)}\,\Theta(A),
    \label{magnetization graphical expansion}
  \end{align}
  where $\mathcal{S}(G)$ is the state space of the worm dynamics, as defined in Section~\ref{Check_worm}, and $Z$ is the partition function of the loop model~\eqref{combinatorial loop model Z}.
  It should therefore come as no surprise that $\<\mathcal{M}^2\>$ can be easily sampled via worm dynamics.

  Indeed, suppose we construct a worm algorithm on $\mathcal{S}(G)$ for which we chose the acceptance probabilities so as to give detailed balance with an arbitrary probability measure
  \begin{equation}
    \mathbb{P}_{\psi}(A,u,v) \propto \psi(A), \qquad (A,u,v)\in\mathcal{S}(G).
  \end{equation}
  If $\<\cdot\>_{\psi}$ denotes expectation with respect to $\mathbb{P}_{\psi}(\cdot)$, then it is obvious that
  \begin{equation}
    \< \indicator_{\mathcal{C}(G)}\>_{\psi} = |V|\,\frac{Z_{\mathcal{C}(G)}(\psi)}{Z_{\mathcal{S}(G)}(\psi)},
    \label{return time identity}
  \end{equation}
  where $Z_{\mathcal{C}(G)}(\psi)$ denotes the sum of $\psi(\cdot)$ over $\mathcal{C}(G)$,
  $Z_{\mathcal{S}(G)}(\psi)$ the sum of $\psi(\cdot)$ over $\mathcal{S}(G)$, and $\indicator_{C(G)}(\cdot)$ denotes the indicator for the event that $(A,u,v)\in\mathcal{S}(G)$ satisfies 
  $A\in\mathcal{C}(G)$, or equivalently that $u=v$. 
  Choosing $\psi(A)=x^{|A|}\,n^{c(A)}\,\Theta(A)$, and combining~\eqref{magnetization graphical expansion} and~\eqref{return time identity} while noting that 
  states in $\mathcal{C}(G)$ do not contain theta graphs, it follows that
  \begin{equation}
    \< \indicator_{\mathcal{C}(G)}\> = \frac{|V|}{\<\mathcal{M}^2\>}
  \end{equation}
  where the expectation on the right is with respect to the $n$-vector spin measure defined by~\eqref{spin model Z} for a given choice of {\em a priori} measure, and the expectation on the 
  left is with respect to the corresponding graphical measure on the worm state space, as inferred from Lemma~\ref{correlation function expansion lemma}.
  The susceptibility of translation invariant systems\footnote{I.e. systems on regular lattices with periodic boundary conditions.} is therefore equal to
  \begin{equation}
    \chi=\frac{\<\mathcal{M}^2\>}{|V|}=\frac{1}{\<\indicator_{\mathcal{C}(G)}\>}.
    \label{return time susceptibility identity}
  \end{equation}

  In fact, the expression~\eqref{return time susceptibility identity} also holds as a relationship between
  the $n\to0$, $p\to1$ limit of Algorithm~\ref{connectivity-checking worm algorithm}, and the susceptibility of the self-avoiding walk problem. 
  Indeed, taking this limit of Algorithm~\ref{connectivity-checking worm algorithm} implies
  \begin{equation}
    \< \indicator_{\mathcal{C}(G)}\> = \frac{1}{\sum_{A\in\{\text{rooted SAWs on $G$}\}}\,x^{|A|}}.
    \label{SAW return time identity}
  \end{equation}
  The denominator on the right-hand side of~\eqref{SAW return time identity} is nothing other than the generating function of variable-length rooted SAWs on $G$, counted according to step length,
  which is precisely the definition of susceptibility for the SAW problem~\cite{MadrasSlade96}. We note that 
  in the $n\to0$, $p\to1$ limit of Algorithm~\ref{connectivity-checking worm algorithm} the quantity $\indicator_{\mathcal{C}(G)}$ is simply the indicator for the $0$-step walk.

  Finally, since $\<\indicator_{\mathcal{C}(G)}\>$ is the stationary probability of the worm chain occupying a state in $\mathcal{C}(G)$,
  the expected time between consecutive visits to $\mathcal{C}(G)$ is simply $\<\mathcal{T}\> = 1/\<\indicator_{\mathcal{C}(G)}\>$, and so for integer $n\ge0$ we have
  \begin{equation}
    \<\mathcal{T}\> = \chi.
  \end{equation}

  \begin{remark}
    For the sake of computational efficiency, the simulations presented in Section~\ref{results} used Algorithm~\ref{coloring worm algorithm} when $n>1$.
    Algorithm~\ref{connectivity-checking worm algorithm} and Algorithm~\ref{coloring worm algorithm} have precisely the same stationary distribution on the subspace $\mathcal{C}(G)$, 
    and therefore the estimates they produce for all loop observables defined on $\mathcal{C}(G)$ must be the same. However, they do give slightly different weights to the non-Eulerian states 
    in $\mathcal{S}(G)$, which implies that $\<\mathcal{T}\>$ will not be precisely equal to the $\chi$ in this case. Nonetheless we expect that the scaling behavior of $\<\mathcal{T}\>$ 
    should still be governed by the magnetic exponent $y_h$. The results for the honeycomb lattice presented in~\cite{LiuDengGaroni11} as well as our results for the hydrogen peroxide lattice
    presented in Section~\ref{results} are consistent with this expectation. 
    In order to test this expectation more systematically, we note that one can in principle construct a generalization of Algorithm~\ref{coloring worm algorithm} in which one introduces the
    vertex colorings to the full worm space $\mathcal{S}(G)$ rather than just the subspace $\mathcal{C}(G)$. 
    This would provide an interesting generalization of the coloring method presented in~\cite{DengGaroniGuoBloteSokal07}. We shall investigate such methods elsewhere.
  \end{remark}

  \section{Simulations}
  \label{results}
  We simulated the loop model~\eqref{combinatorial loop model Z} on 
  finite hydrogen-peroxide lattices with periodic boundary conditions, for loop fugacities $n=0,0.5,1,1.5,2,3,4,5$ and $10$.
  For $n=0,0.5,1$ we used Algorithm~\ref{connectivity-checking worm algorithm}, with $p$ set to $1$ for $n=0$ and to $1/2$ for $n=0.5,1$.
  For $n>1$ we used Algorithm~\ref{coloring worm algorithm}, so as to avoid the need for non-local connectivity queries.
  
  For each value of $n$, we ran simulations of several $L\times L\times L$
  systems at several values of $x$. The length unit is chosen as one half of 
  the 8-site hydrogen-peroxide cell, so that the systems contain $L^3$ sites.
  The lattice sizes $L$ were taken in the range 8 to 128.
  Additional simulations for $L=256$ were performed at the critical point $x_c$ as estimated from the smaller system sizes.
  For each choice of $(n,x,L)$ we performed multiple independent simulations, as a means of implementing (trivial) parallelization.
  For each independent simulation, statistical errors were estimated by partitioning the data into 1000 bins, and computing error bars using blocking.
  The resulting means and error bars of the independent runs were then combined to produce the final estimates for each choice of $(n,x,L)$.
  For $n\neq0$, we performed between 100 and 300 independent simulations for each choice of $(n,x,L)$, each simulation consisting of $10^6$ returns to the Eulerian subspace, 
  the first $10^4$ returns being discarded as burn-in.
  For $n=0$, we performed 50 independent simulations for each choice of $(x,L)$, each simulation consisting of between 
  $0.5 \times 10^6\,L^3$ and $1.25 \times 10^6\,L^3$ iterations (hits), with the first $10^3\,L$ iterations being discarded as burn-in.
  For the $n=0$ case we sampled observables every $L^3/4$ iterations, since we found in all cases that taking lags of this size guaranteed the autocorrelation function for $\mathcal{N}_b$ was smaller than
  $0.1$. (See Section~\ref{dynamic results} for relevant definitions concerning autocorrelations.)

  For each of the quantities discussed in Section~\ref{observables} we performed a least-squares fit of our Monte Carlo data to an appropriate finite-size scaling (FSS) expression.
  As a precaution against excessive corrections to scaling, we imposed a lower cutoff $L \geq L_{\rm min}$ on the data points admitted to the fit,
  and we studied systematically the effects on the fit due to variations of the value of $L_{\rm min}$.
  The error bars that we report for our fits are composed of the statistical error (one standard error) plus a subjective estimate of the systematic error.
  To estimate the systematic error we compared multiple distinct fits for each quantity, corresponding to different choices of correction terms included in the FSS formulas,
  and/or different choices of $L_{\min}$.
  Given that the FSS formulas are nonlinear, we used the Levenberg-Marquardt algorithm to perform the least-squares fits.

  \subsection{The wrapping probability $R$}
  According to finite-size scaling theory, the expected behavior of the dimensionless wrapping probability $R(x,L)$ as a function of the linear system size $L$ and temperature-like parameter $x$ is
  \begin{equation}
    \begin{split}
      \label{R ansatz}
      R &= R^{(0)}+a_1(x-x_c)L^{y_t}+a_2(x-x_c)^2L^{2y_t}+a_3(x-x_c)^3L^{3y_t}+ a_4(x-x_c)^4L^{4y_t} \\
      &\quad+ c(x-x_c)^2L^{y_t}+b_1L^{y_1}+b_2L^{y_2}+b_3L^{y_3} + e(x-x_c)L^{y_t+y_1}+\ldots
    \end{split}
  \end{equation}
  where $x_c$ denotes the critical point, $y_t$ the thermal exponent, and $y_1,y_2,y_3$ are negative correction-to-scaling exponents.
  
  We fitted Eq.~(\ref{R ansatz}) to our Monte Carlo data, and the resulting estimates for $x_c$ and $y_t$ are reported in Table \ref{R fits}.
  In each fit the values of the correction-to-scaling exponents were fixed, while $x_c$, $y_t$ were free parameters. 
  For $n=1$, the correction-to-scaling exponents were fixed to $y_1=-0.815$, $y_2=-1.963$, and $y_3=-3.375$ \cite{DengBlote03}, while for all other $n>0$ we set $y_1=-0.85$, $y_2=-1.8$ and $y_3=-3$. 
  The values of $y_1$ and $y_2$ are based on estimates found in~\cite{NewmanRiedel84} for $n=0,1,2,3$ and the supposedly weak dependence of these exponents on $n$.
  
  For $n\le3$, each of the coefficients $a_1,\ldots,a_4,b_1,c$ were free parameters in the fits, while the coefficient $e$ was set identically to zero.
  As a means of gauging systematic errors, we performed fits with $b_2,b_3$ set identically to zero, as well as fits in which $b_2,b_3$ were free parameters,
  and compared the resulting estimates for $y_t$ and $x_c$.
  For each $n\le3$, our final estimates of $y_t$ and $x_c$ and their error bars were obtained by comparing these two fits.
  For $n=4,5,10$, the strength of the corrections to scaling demanded that we include all terms in~\eqref{R ansatz} in our fits, 
  so that each of the coefficients $a_1,\ldots,a_4,b_1,\ldots,b_3,c,e$ were free parameters in these fits.

  \begin{figure}
    \centering
    \subfigure[$R$ versus $x$ for $n=1$.]{\label{R figure}
      \includegraphics[angle=270,scale=0.25]{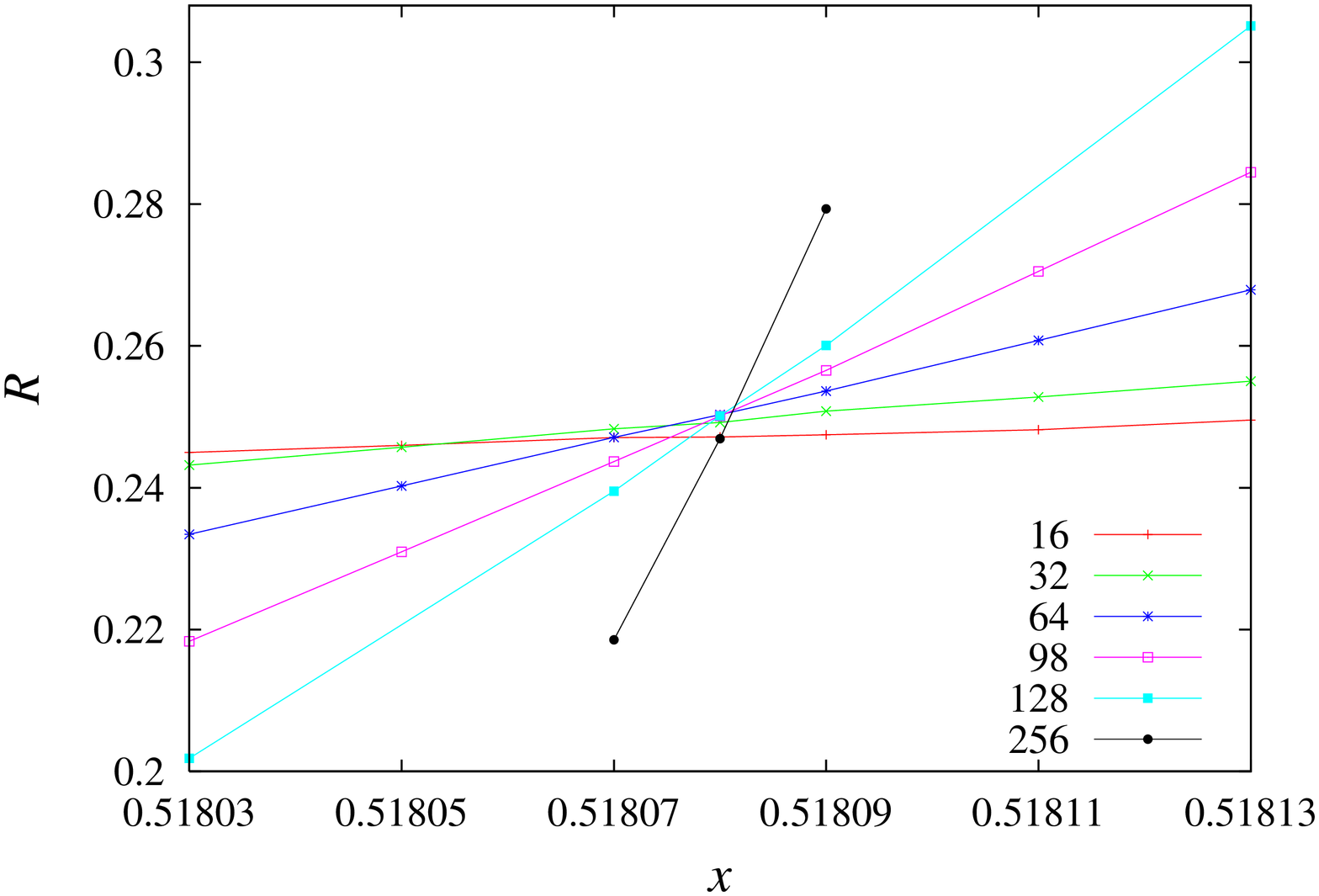}
    }
    \subfigure[$Q_b$ versus $x$ for $n=0$.]{\label{SAW Qb figure}
      \includegraphics[angle=270,scale=0.25]{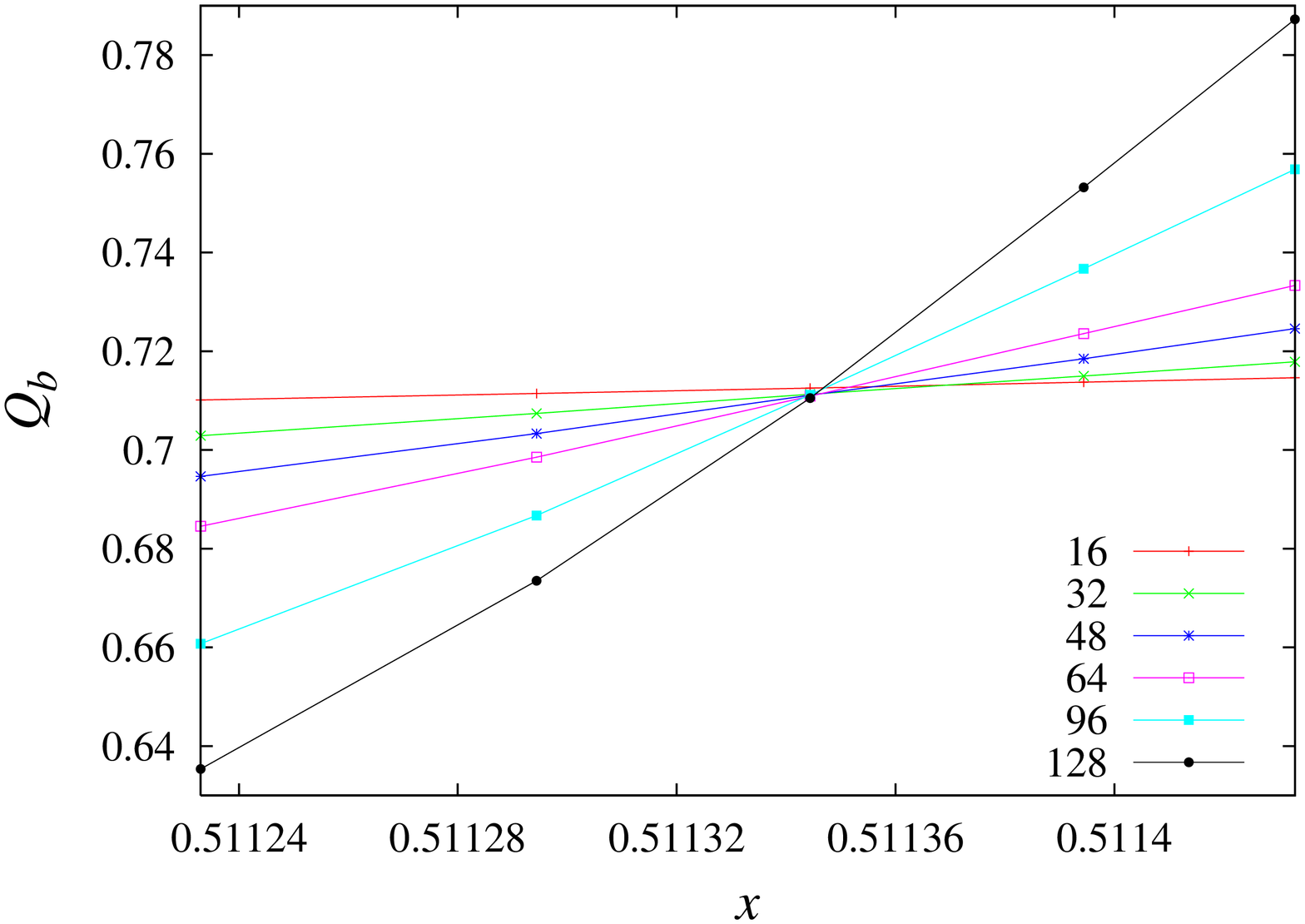}
    }
    \caption{\label{Binder figures} Wrapping probability $R$ for the $n=1$ loop model and dimensionless ratio $Q_b$ for self-avoiding walks ($n=0$),
      versus the bond (step-length) fugacity $x$, for linear system sizes 16 to 256.}
  \end{figure}

  \begin{table}
    \centering
    \caption{\label{R fits}Estimated critical point $x_c$ and thermal exponent $y_t$ for a number of values of the loop fugacity $n$, 
      as determined from least-squares fits of the wrapping probability $R$.
    }
    \medskip
        {\footnotesize
          \begin{tabular}{|l|l|l|l|l|l|}
            \hline
            $n$      & $x_c$        & $y_t$    & $R^{(0)}$ & $a_1$    & $b_1$ \\
            \hline
            $0.5$    & 0.5146078(6) & 1.651(5) & 0.1467(2) & 1.02(3)  & $-$0.08(3)  \\
            $1$      & 0.5180815(3) & 1.588(2) & 0.2524(3) & 1.41(1)  & $-$0.031(5) \\
            $1.5$    & 0.5217878(5) & 1.539(4) & 0.3326(5) & 1.47(3)  & $-$0.062(5) \\
            $2$      & 0.5257533(5) & 1.489(3) & 0.3990(4) & 1.48(1)  & $-$0.077(6) \\
            $3$      & 0.5345904(8) & 1.398(3) & 0.5020(6) & 1.33(1)  & $-$0.144(5) \\
            $4$      & 0.544904(2)  & 1.329(8) & 0.5861(6) & 1.08(2)  & $-$0.18(2) \\
            $5$      & 0.557096(5)  & 1.275(12)& 0.6548(6) & 0.82(3)  & $-$0.21(2)  \\
            $10$     & 0.67367(2)   & 1.142(15)& 0.8638(5) & 0.12(2)  & $-$0.23(2)  \\
            \hline
          \end{tabular}
        }
  \end{table}

  \subsection{The dimensionless ratios $Q_l$ and $Q_b$}
  As a consistency check on the values of $x_c$ and $y_t$ determined from $R$, we performed analogous fits of $Q_l$, again on the basis of~\eqref{R ansatz},
  with the same procedure for fixing the correction exponents and coefficient values.
  The resulting estimates for $x_c$ and $y_t$ are reported in Table \ref{Q fits}. The agreement with the corresponding estimates produced by the $R$ fits is clearly excellent.

  \begin{table}
    \centering
    \caption{\label{Q fits}Estimated critical point $x_c$ and thermal exponent $y_t$ for a number of values of the loop fugacity $n$, 
      as determined from least-squares fits of the dimensionless ratios $Q_l$ and $Q_b$.
    }
    \medskip
        {\footnotesize
          \begin{tabular}{|l|l|l|l|l|l|}
            \hline
            $n$   & $x_c$        & $y_t$    &  $Q^{(0)}$ & $a_1$      & $b_1$  \\
            \hline
            $0$   & 0.5113445(2) & 1.701(2) & 0.7017(2)  & 0.675(4) & 0.11(2)  \\
            $0.5$ & 0.5146084(7) & 1.653(5) & 0.4858(6)  & 0.578(6) & 0.012(6) \\
            $1$   & 0.5180813(4) & 1.586(7) & 0.6158(4)  & 0.836(5) & 0.042(5) \\
            $1.5$ & 0.5217884(7) & 1.532(8) & 0.6884(5)  & 0.750(6) & 0.052(5) \\
            $2$   & 0.5257539(7) & 1.487(3) & 0.7316(4)  & 0.644(6) & 0.066(4) \\
            $3$   & 0.534595(4)  & 1.398(4) & 0.7827(3)  & 0.435(5) & 0.073(3) \\
            $4$   & 0.544907(4)  & 1.332(7) & 0.8086(4)  & 0.270(6) & 0.080(7) \\
            $5$   & 0.557108(8)  & 1.280(10)& 0.8252(4)  & 0.162(5) & 0.12(2)  \\
            $10$  & 0.67373(7)   & 1.145(30)& 0.8573(3)  & 0.016(5) & 0.10(2)  \\
            \hline
          \end{tabular}
        }
  \end{table}

  The results for $R$ and $Q_l$ applied only to $n>0$, however analogous fits can be performed for $n=0$ using $Q_b$. 
  Following the same procedure as for $R$ and $Q_l$ we thereby obtained the estimated values of $x_c$ and $y_t$ for $n=0$ presented in Table~\ref{Q fits}.
  While this treatment of the SAW case appears natural within the context of the general loop model, we note that it may be considered somewhat unusual to deliberately study SAWs on a finite system.
  Indeed, one reason why SAWs are often considered easier to study via simulation than other models in statistical mechanics is precisely because it is possible to study SAWs on an infinite lattice.
  One might therefore be led to expect that $Q_b$ will be a rather poor estimator for the critical point. However, it is clear from Fig.~\ref{SAW Qb figure} that this is not the case,
  and in fact the sharpness of the point of intersection of the $Q_b$ curves in Fig.~\ref{SAW Qb figure} is quite remarkable. It would be interesting to see if such behavior holds on other lattices,
  such as the simple-cubic lattice.

  \subsection{Mean-square loop length}
  The finite-size scaling behavior of $\<\mathcal{L}_2\>$ is expected to be~\cite{LiuDengGaroni11}
  \begin{equation}
    \<\mathcal{L}_2\> = L^{2y_l-d}(a_0+a_1(x-x_c)L^{y_t}+b_1L^{y_1}+b_2L^{y_2}+b_3L^{y_3}+\ldots)
    \label{L2 ansatz}
  \end{equation}
  where $y_l$ is the fractal dimension characterizing loop length.
  We fitted Eq.~(\ref{L2 ansatz}) to our Monte Carlo data for $\<\mathcal{L}_2\>$, fixing $y_t$ and $x_c$ to their values as reported in Table \ref{R fits}.
  The corrections-to-scaling exponents were fixed according to the procedure described for $R$.
  And again, as a means of gauging systematic errors, for $n\le 3$ we performed fits with $b_2,b_3$ set identically to zero, as well as fits in which $b_2,b_3$ were free parameters,
  and compared the resulting estimates for $y_t$ and $x_c$.
  The resulting estimates for $y_l$ are reported in Table \ref{loop and magnetic exponents}.

  \begin{table}
    \centering
    \caption{\label{loop and magnetic exponents}Estimated values of the scaling exponents $y_l$ and $y_h$, for a number of values of the loop fugacity $n$, 
      as determined from least-squares fits of $\<\mathcal{L}_2\>$ and $\<\mathcal{T}\>$. 
      For $n\le1$, Algorithm~\ref{connectivity-checking worm algorithm} was used, while Algorithm~\ref{coloring worm algorithm} was used for $n>1$.
    }
    \medskip
        {\footnotesize
	  \begin{tabular}{|l|ll|lll|}
	    \hline
 	    \multicolumn{1}{|c}{} & \multicolumn{2}{c|}{Fits for $\<\mathcal{L}_2\>$} & \multicolumn{3}{c|}{Fits for $\<\mathcal{T}\>$} \\
 	    \hline
 	    $n$    & $y_l$      & $a_0$         & $y_h$          & $a_0$      & $a_1$ \\
 	    \hline
 	    $0$    & \---       & \---          & 2.4875(7) & 0.796(5) & $-$6.257(2) \\
 	    $0.5$  & 1.723(3)   &  0.64(2)      & 2.482(4)  & 1.11(2)  & $\phantom{-}$6.8(1)   \\
 	    $1$    & 1.734(4)   &  1.16(3)      & 2.483(3)  & 0.97(2)  & $\phantom{-}$5.04(6)  \\
 	    $1.5$  & 1.755(3)   &  1.44(3)      & 2.482(3)  & 0.91(2)  & $\phantom{-}$3.80(5)  \\
 	    $2$    & 1.765(3)   &  1.80(2)      & 2.483(2)  & 0.86(2)  & $\phantom{-}$3.16(3)  \\
 	    $3$    & 1.795(3)   &  2.18(2)      & 2.482(2)  & 0.80(2)  & $\phantom{-}$2.32(4)  \\
 	    $4$    & 1.816(3)   &  2.62(2)      & 2.483(2)  & 0.77(2)  & $\phantom{-}$1.63(4)  \\
 	    $5$    & 1.834(6)   &  2.9(1)       & 2.483(3)  & 0.74(2)  & $\phantom{-}$1.22(4)  \\
 	    $10$   & 1.901(8)   &  5.6(2)       & 2.486(3)  & 0.75(4)  & $\phantom{-}$0.26(5)  \\
	    \hline
	  \end{tabular}
        }
  \end{table}

  \subsection{Susceptibility $\chi$ - average worm return time}
  The finite-size scaling behavior of $\<\mathcal{T}\>$ is expected to be~\cite{LiuDengGaroni11}
  \begin{equation}
    \<\mathcal{T}\>=L^{2y_h-d}(a_0+a_1(x-x_c)L^{y_t}+b_1L^{y_1}+\ldots)
    \label{T FSS ansatz}
  \end{equation}
  where $y_h$ is the magnetic exponent.
  We fitted Eq.~(\ref{T FSS ansatz}) to our Monte Carlo data for $\<\mathcal{T}\>$, fixing $y_t$ and $x_c$ to their values as reported in Table~\ref{R fits} (Table~\ref{Q fits} for $n=0$).
  For the fits to \eqref{T FSS ansatz} it was found sufficient to include only one correction-to-scaling term, and $y_1$ was fixed using the same procedure as for the $R$ fits.
  The resulting estimates for $y_h$ are reported in Table \ref{loop and magnetic exponents}.

  A remark concerning our use of Algorithm~\eqref{coloring worm algorithm} is in order.
  For $n\le1$, the values of $\<\mathcal{T}\>$ were computed using Algorithm~\ref{connectivity-checking worm algorithm}, and so the discussion of Section~\ref{susceptibility expansion} implies that
  for $n=0,1$ the mean return time $\<\mathcal{T}\>$ is precisely equal to the susceptibility of the SAW and Ising models respectively.
  For $n>1$ however, we used Algorithm~\ref{coloring worm algorithm} in our simulations, and so the discussion in Section~\ref{susceptibility expansion} does not imply the exact equality
  of $\<\mathcal{T}\>$ and $\chi$ in these cases. However, it was found empirically in~\cite{LiuDengGaroni11} that on the honeycomb lattice while $\<\mathcal{T}\>$ may not be precisely equal to 
  $\chi$ when using Algorithm~\ref{coloring worm algorithm}, the two quantities scale in the same way at criticality, so measuring $\<\mathcal{T}\>$ in Algorithm~\ref{coloring worm algorithm}
  was indeed found to be an accurate method for computing $y_h$. 
  As we discuss further in Section~\ref{discussion}, our estimates for $y_h$ on the hydrogen peroxide lattice also agree within error bars with previously-obtained results for the O($n$) universality class.

  \subsection{Dynamic behavior}
  \label{dynamic results}
  When $n=1$, Algorithms~\ref{connectivity-checking worm algorithm} and~\ref{coloring worm algorithm} coincide, and in this case a careful study of the dynamic critical behavior was carried out 
  in~\cite{DengGaroniSokal07c}. In particular, it was found that in three dimensions the Li-Sokal bound (see below) appeared to be sharp. 
  In order to explore the efficiency of the worm algorithms for more general $n$, we studied the dynamic critical behavior of both Algorithm~\ref{connectivity-checking worm algorithm} and 
  Algorithm~\ref{coloring worm algorithm} for the case $n=2$.
  For comparison, we also studied the dynamic critical behavior of a local plaquette update algorithm.

  Consider an observable $\mathcal{O}$. A realization of the worm Markov chain gives rise to a time series $\mathcal{O}(t)$.
  The autocorrelation function of $\mathcal{O}$ is defined to be~\cite{SokalLectures}
  \begin{equation*}
    \rho_{\mathcal{O}}(t) = \frac{\< \mathcal{O}(0)\mathcal{O}(t)\> - \< \mathcal{O}\>^2}{\var(\mathcal{O})},
  \end{equation*}
  where $\< \cdot \>$ denotes expectation with respect to the stationary distribution.
  From $\rho_{\mathcal{O}}(t)$ we define the integrated autocorrelation time as
  \begin{equation}
    \tau_{{\rm int},\mathcal{O}} \equiv  \frac{1}{2}\,\sum_{t = -\infty}^{\infty}\rho_{\mathcal{O}}(t),
  \end{equation}
  and the exponential autocorrelation time as
  \begin{equation}
    \tau_{{\rm exp},\mathcal{O}}  \equiv \limsup_{t \to \pm\infty} \frac{|t|}{- \log |\rho_{\mathcal{O}}(t)|}.
  \end{equation}
  Finally, the exponential autocorrelation time of the system is defined as 
  \begin{equation}
    \tau_{\exp} = \sup_{\mathcal{O}}\,\tau_{\exp,\mathcal{O}}
  \end{equation}
  where the supremum is taken over all observables $\mathcal{O}$. 
  This autocorrelation time therefore samples the relaxation rate of the slowest mode of the system. 
  All observables that are not orthogonal to this slowest mode satisfy $\tau_{\exp,\mathcal{O}}=\tau_{\exp}$.

  The autocorrelation times typically diverge as a critical point is approached, most often like $\tau \sim \xi^z$, where $\xi$ is the spatial correlation length and $z$ is a dynamic exponent.
  This phenomenon is referred to as {\rm critical slowing-down}~\cite{SokalLectures}.
  In the case of a finite lattice at criticality, we define the dynamic critical exponents $z_{{\rm int},\mathcal{O}}$, $z_{\exp,\mathcal{O}}$ and $z_{\rm exp}$ by
  \begin{equation}
    \begin{split}
      \label{tau definitions}
      \tau_{{\rm int},\mathcal{O}} \sim L^{z_{{\rm int},\mathcal{O}}}, \qquad \tau_{{\rm exp},\mathcal{O}} \sim L^{z_{{\rm exp},\mathcal{O}}} \qquad \tau_{\rm exp} \sim L^{z_{\rm exp}}.
    \end{split}
  \end{equation}

  Note that when defining $z_{\exp}$ and $z_{\text{int},\mathcal{O}}$ it is natural to express time in units of {\em sweeps} of the lattice, i.e. $L^d$ hits.
  However, when using Algorithms~\ref{connectivity-checking worm algorithm} or~\ref{coloring worm algorithm}, 
  observables are sampled only when the chain visits the Eulerian subspace.
  Given that $\<\mathcal{T}\>$ scales like the susceptibility, this occurs roughly every $L^{d-2X_h}$ iterations, where $X_h$ is the magnetic scaling dimension.
  Since one sweep therefore takes of order $L^{2X_h}$ visits to $\mathcal{C}(G)$, in units of ``visits to $\mathcal{C}(G)$'' we have $\tau\sim L^{z+2X_h}$.
  Consequently, we fit our dynamic data to the ansatz
  \be
  \tau_{{\rm int},\mathcal{O}}=a+bL^{z_{{\rm int},\mathcal{O}}+2X_h}
  \ee
  to obtain the exponent $z_{{\rm int},\mathcal{O}}$.

\begin{figure}[ht]
    \begin{center}
      \includegraphics[angle=270,scale=0.4]{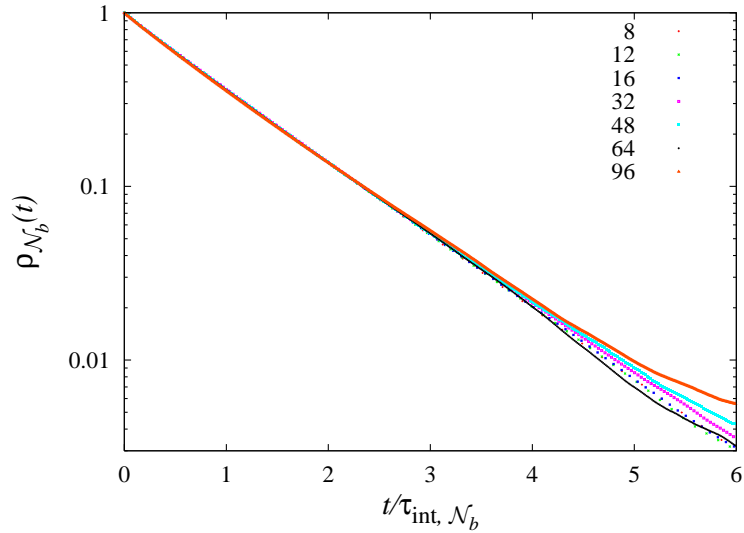}
      \caption{ Autocorrelation function $\rho_{\mathcal{N}_b}(t)$ versus $t/\tau_{{\rm int},\mathcal{N}_b}$.
        The almost-pure exponential decay of $\rho_{\mathcal{N}_b}(t)$ suggests $z_{\rm exp}\thickapprox z_{{\rm int},\mathcal{N}_b}$.
      }
      \label{rho}
      \vspace{-0.7cm}
    \end{center}
  \end{figure}

  In this work, we determined $\rho_{\mathcal{O}}(t)$ and $\tau_{{\rm int},\mathcal{O}}$ for the observables $\mathcal{N}_b$, $\mathcal{L}_2$, and $\mathcal{N}_l$.
  This was done for $n=2$, using both Algorithm~\ref{connectivity-checking worm algorithm} and Algorithm~\ref{coloring worm algorithm}.
  We find that the slowest of these observable is $\mathcal{N}_b$.
  Figure \ref{rho} shows $\rho_{\mathcal{N}_b}$ as a function of $t/\tau_{{\rm int},\mathcal{N}_b}$, 
  where a nearly pure exponential decay is observed, suggesting $z_{\rm exp}\thickapprox z_{{\rm int},\mathcal{N}_b}$.

  For Algorithm~\ref{connectivity-checking worm algorithm} we find $z_{{\rm int},\mathcal{N}_b}=0.19~(3)$, 
  while for Algorithm~\ref{coloring worm algorithm} we find $z_{{\rm int},\mathcal{N}_b}=0.243~(12)$.
  It is perhaps unsurprising that the bonds should decorrelate faster when using Algorithm~\ref{connectivity-checking worm algorithm} compared with Algorithm~\ref{coloring worm algorithm},
  since in the latter case a non-trivial subset of the bonds is always being held fixed at any given time.
  We emphasize, however, that these values of $z_{{\rm int},\mathcal{N}_b}$ are inherent properties of the underlying Markov chains, and do not encode any information regarding the amount of CPU time required 
  in practice to execute each MC update.
  In particular, despite the fact that Algorithm~\ref{connectivity-checking worm algorithm} suffers from slightly weaker critical slowing down, in practice
  Algorithm~\ref{coloring worm algorithm} actually outperforms our implementation of Algorithm~\ref{connectivity-checking worm algorithm}, because it avoids the need for costly non-local
  connectivity queries. 

  For comparison, we also determined $z_{\text{int},\mathcal{N}_b}$ for a simple plaquette-update algorithm. Like worm algorithms, these plaquette update algorithms can be used for 
  arbitrary real $n>0$. However, a rough estimation for $n=2$ yields $z_{{\rm int},\mathcal{N}_b}=2.26~(5)$, which is significantly larger than for the worm algorithm.
  Similar values of $z$ for plaquette algorithms have been observed previously~\cite{WinterJankeSchakel08}.
  In addition, we note that the performance of the worm algorithm compares very favorably with Wolff's embedding algorithm~\cite{Wolff89}, 
  which is perhaps the most widely-used Monte Carlo method for studying O($n$) spin models, and which was found in~\cite{HasenbuschMeyer90} to have $z_{{\rm int},\text{energy}}=0.46(3)$ when $(n,d)=(2,3)$.

  Finally, it is interesting to compare these results with the $n=1$ case~\cite{DengGaroniSokal07c} where it was found that $z_{\text{int},\mathcal{N}_b}\approx\alpha/\nu\approx0.17$. This implies that 
  the Li-Sokal bound~\cite{LiSokal89} $z_{\text{int},\mathcal{N}_b}\ge\alpha/\nu$, which can be easily generalized to worm algorithms, appears to be sharp for $n=1$.
  By contrast, for $n=2$ we have $\alpha/\nu\approx-0.026$, so that the Li-Sokal bound does not appear to be close to sharp in this case.

  \section{Discussion}
  \label{discussion}
  Using the worm algorithms introduced in~\cite{LiuDengGaroni11}, we have studied the critical behavior of the loop model~\eqref{combinatorial loop model Z} on a three-dimensional 3-regular lattice, 
  the hydrogen peroxide lattice, for a range of integer and non-integer loop fugacities $0\le n\le10$.
  To our knowledge, this is the first direct Monte Carlo study of loop models with general $n$ in three dimensions.
  A study of the dynamic critical behavior of these algorithms for $n=2$ shows that the critical-slowing down is only very weak; 
  not only is it much weaker than for local plaquette-update algorithms, but also significantly weaker than for the Wolff embedding algorithm~\cite{Wolff89,HasenbuschMeyer90}.
  Combined with the previous study of the $n=1$ case~\cite{DengGaroniSokal07c}, these results strongly suggest that worm algorithms provide a very effective method for studying loop models in three dimensions.

  Our simulations show that the loop model~\eqref{loop model Z} undergoes a continuous phase transition, with critical exponents that depend on the loop fugacity $n$.
  Our best estimates of the exponents $y_t$ and $y_h$ are reported in Table~\ref{exponent comparisons}.
  For integer $n$, these exponent estimates are consistent with the O($n$) universality class; for comparison, we list in Table~\ref{exponent comparisons} a number of relevant
  exponent estimates from the literature.
  For the cases $n=0$ and $n=1$ this is entirely to be expected, since exact mappings unambiguously identify the loop model with the SAW and Ising models respectively,
  and these mappings are valid in a range of $x$ which includes the critical point $x_c$. 
  In fact, for the $n=0$ case Algorithm~\ref{connectivity-checking worm algorithm} actually directly simulates the grand canonical ensemble of SAWs.
  More generally, one might expect from the identity~\eqref{spin model Z} that the loop model should in fact be in the same universality class as the $n$-vector model for any integer $n$.
  However, for $n>1$ the situation is slightly more subtle, for two reasons.
  Firstly, the spin model to which the loop model maps has positive weights only for $x<1/n$, and for $n\ge2$ we find empirically that $x_c$ lies outside this range.
  Secondly, and perhaps more importantly, the mapping from the spin model~\eqref{spin model Z} to the loop model~\eqref{loop model Z} is many-valued;
  it maps the loop model~\eqref{loop model Z} to a number of distinct spin models, whose Hamiltonians posses different
  symmetry groups and which therefore might be expected to belong to distinct universality classes.

  \begin{table}
    \centering
    \caption{\label{exponent comparisons} Comparison of the estimated loop-model exponents $y_t$ and $y_h$ with existing literature.
      For $n=0$ we compare with known SAW exponents, while for integer $n\ge1$ we compare with the corresponding exponents of the O($n$) spin model.
      No error bounds were given in~\cite{AntonenkoSokolov95} for their estimate of $y_t$ at $n=10$.
    }
    \medskip
        {\footnotesize
          \begin{tabular}{|l|l|l|l|l|}
            \hline
            \multicolumn{1}{|c}{} & \multicolumn{2}{c|}{$y_t$} & \multicolumn{2}{c|}{$y_h$}\\
            \hline
            $n$    & This work & Literature & This work     & Literature \\
            \hline
            $0$    & 1.701(2)  & 1.7018(6)\cite{PelisettoVicari07}, 1.70179(5)\cite{HsuNadlerGrassberger04}, 1.70185(2)\cite{Clisby10} & 2.4875(7)  & 2.4849(5)\cite{CaraccioloCausoPelisetto98} \\
            $0.5$  & 1.653(5)  & \---      & 2.482(4) & \--- \\
            $1$    & 1.588(2)  & 1.5868(3)\cite{DengBlote03} & 2.483(3) & 2.48180(8)\cite{CampostriniPelissettoRossiVicari02}, 2.4816(1)\cite{DengBlote03}\\
            $1.5$  & 1.538(4)  & \---      & 2.482(3) & \--- \\
            $2$    & 1.488(3)  & 1.4888(2)\cite{CampostriniHasenbuschPelissettoVicari06}& 2.483(2) & 2.4810(1)\cite{CampostriniHasenbuschPelissettoVicari06} \\
            $3$    & 1.398(2)  & 1.406(10)\cite{CampostriniHasenbuschPelissettoRossiVicari02}, 1.405(2)\cite{HasenbuschVicari11} & 2.482(2) & 2.4830(5)\cite{CampostriniHasenbuschPelissettoRossiVicari02},
            2.4811(15)\cite{HasenbuschVicari11}\\
            $4$    & 1.332 (7) & 1.3375(15)\cite{Deng06}, 1.333(4)\cite{HasenbuschVicari11} & 2.483(2) & 2.4820(2)\cite{Deng06}, 2.4820(15)\cite{HasenbuschVicari11}\\
            $5$    & 1.275(12) & 1.309(7)\cite{ButtiToldin05} & 2.483(3) & 2.4845(15)\cite{ButtiToldin05}\\
            $10$   & 1.142(15)  & 1.164(\---)\cite{AntonenkoSokolov95} & 2.486(3) & 2.488(1)\cite{AntonenkoSokolov95}\\
            \hline
          \end{tabular}
        }
  \end{table}
  To put these observations in context, let us briefly recount the relationship between the O($n$) model and the two discrete cubic models discussed in Section~\ref{susceptibility expansion}.
  Renormalization group arguments~\cite{PelissettoVicari02} predict the following scenario.
  There exists a critical $n_c$, believed to be $\approx3$ in three dimensions, below which the cubic models and O($n$) models share the same universality class, 
  but beyond which they do not. For $n>n_c$, the face-cubic model is believed to undergo a first order transition, while the critical behavior of the corner-cubic model is believed to 
  be governed by a distinct {\em cubic} fixed point. The discrepancies between the predicted exponent values of the O($n$) and cubic fixed points at $n=3$ are too small to detect using current
  approximations, however they increase with $n$.
  Indeed, as $n\to\infty$,  the O($n$) model approaches the spherical model~\cite{BerlinKac52,Stanley68}, with exponents $y_{t, \text{spherical}}=1$ and $y_{h, \text{spherical}}=5/2$, while
  the corner-cubic model can be reinterpreted as a constrained Ising model~\cite{Aharony76} so that its exponents are given by a Fisher renormalization of the Ising exponents,
  $y_{t, \text{cubic}} = d - y_{t, \text{Ising}} = 1.4132(3)$ and, $y_{h, \text{cubic}}=y_{h, \text{Ising}} = 2.4816(1)$.
  For finite $n$ these exponents have corrections of $O(1/n)$; for example~\cite{CarmonaPelissettoVicari00} predicts $y_{t, \text{cubic}}=1.416(12)$, $1.401(16)$, $1.404(12)$ for $n=3,4,8$.

  Our estimates of $y_t$ reported in Table~\ref{exponent comparisons} agree within error bars with the corresponding O($n$) values from the literature when $n\le4$.
  For $n=5,10$ the agreement is not perfect, but still quite convincing. Furthermore, the $n$-dependence of our $y_t$ estimates is entirely consistent with the limiting O($n$) value of $y_t=1$ as $n\to\infty$.
  The behavior of $y_h$ in Table~\ref{exponent comparisons}, while only weakly dependent on $n$, is also consistent with the O($n$) universality class.
  By contrast, our $y_t$ estimates are entirely inconsistent with the cubic fixed point.
  Based on these observations it seems reasonable to conclude that the critical behavior of the loop model~\eqref{loop model Z} on the hydrogen peroxide lattice belongs to the O($n$) universality class.
  
  Finally, we note that the literature also contains results for the loop exponent $y_l$ in three dimensions.
  Recently, Winter et al.~\cite{WinterJankeSchakel08} determined the fractal dimension of the high-temperature graphs at criticality for the Ising and the XY model on a cubic lattice.
  Using a plaquette update algorithm, they estimated the fractal dimensions of the Ising and the XY model as $D=1.7349~(65)$ and $D=1.7626~(66)$ respectively.
  Furthermore, Prokof'ev and Svistunov \cite{ProkofevSvistunov06} reported the fractal dimension of the graph expansion of the complex $|\phi|^4$ theory at its critical point as $D=1.7655~(20)$.
  From the expected equivalence in universality we may compare this results with our result of $y_l$ for the $n=2$ model.
  As can be seen from the Table \ref{loop and magnetic exponents}, our estimates of $y_l$ are in good agreement with the cited literature.

  \section*{Acknowledgments}
  This work was supported in part by the National Nature Science Foundation of China under Grant No. 10975127, 
  the Specialized Research Fund for the Doctoral Program of Higher Education under Grant No. 20113402110040,
  and the Chinese Academy of Sciences. It was also supported under the Australian Research Council's Discovery Projects funding scheme (project number DP110101141),
  and T.G. is the recipient of an Australian Research Council Future Fellowship (project number FT100100494). 
  H.B. thanks the Lorentz Fund for financial support, and the University of
  Science and Technology of China in Hefei for hospitality extended to him.

  \bibliographystyle{elsarticle-num}

  \bibliographystyle{elsarticle-num}

\end{document}